\documentclass[onecollarge,referee]{svjour2}

\usepackage{epsfig}
\usepackage{times}

\begin{document}

\title{A mathematical analysis of the evolution of perturbations in a 
modified Chaplygin gas model}

\author{Sandro Silva e Costa \and Maximiliano Ujevic \and Alesandro Ferreira dos Santos}

\institute{S.S. Costa \and M. Ujevic \at Centro de Ci\^encias Naturais e 
Humanas, Universidade Federal do ABC, 09210-170, Santo Andr\'e, S\~ao 
Paulo, Brasil \\\email{sandro.costa@ufabc.edu.br}, \and M. Ujevic 
\\\email{maximiliano.tonino@ufabc.edu.br}, \\ A.F. dos Santos \at 
Departamento de F\'isica, Universidade Federal de Mato Grosso, 
78060-900, Cuiab\'a, Mato Grosso, Brasil 
\\\email{alesandroferreira@fisica.ufmt.br}}


\maketitle

\begin{abstract} 
One approach in modern cosmology consists in supposing that dark matter 
and dark energy are different manifestations of a single 
`quartessential' fluid. Following such idea, this work presents a study 
of the evolution of perturbations of density in a flat cosmological 
model with a modified Chaplygin gas acting as a single component. Our 
goal is to obtain properties of the model which can be used to 
distinguish it from another cosmological models which have the same 
solutions for the general evolution of the scale factor of the universe, 
without the construction of the power spectrum. Our analytical results, 
which alone can be used to uniquely characterize the specific model 
studied in our work, show that the evolution of the density contrast can 
be seen, at least in one particular case, as composed by a spheroidal wave function. We also present a numerical analysis which clearly indicates as one interesting feature of 
the model the appearence of peaks in the evolution of the density 
constrast.
\end{abstract}


\maketitle

\section{Introduction}

In an expanding universe distances grow following the expression
\begin{equation}
r\left(t\right)=r_0a\left(t\right)\,,
\end{equation}
where $a\left(t\right)$ is the {\it scale factor}, which indicates how a 
distance $r$ varies with time, with $r_0$ being the present value of $r$. A 
Taylor's expansion of such expression around the present time $t_0$ 
produces the approximate relation
\begin{equation}
a\left(t\right)\approx
1+H_0\left(t-t_0\right)-\frac{1}{2}q_0H_0^2\left(t-t_0\right)^2\,,
\end{equation}
where $a\left(t_0\right)=1$, 
$H_0=\left(\dot{a}/a\right)\left|_{t=t_0}\right.$ and 
$q_0=-\left(\ddot{a}a/\dot{a}^2\right)\left|_{t=t_0}\right.$, with 
$\dot{a}$ meaning the time derivative of the function $a$. There are 
several ways of measuring the quantities $H_0$ (called Hubble constant) 
and $q_0$ (the decceleration parameter). Presently, the most cited 
method consists of the observation of supernovae in distant galaxies 
\cite{supernovae1,supernovae2}, and its use brought to light the idea that the 
universe is passing through a phase of accelerated expansion. So, a 
major problem confronting cosmologists today may be resumed in the 
question ``what causes the acceleration of the universe's expansion?'' 
Since there are yet no definitive and compelling answers, such issue, 
named as the problem of dark energy, remains open to debate.

Another parallel puzzle of modern cosmology involves the contradiction 
between the observed movements of stars in the periphery of galaxies 
(and of galaxies in clusters of galaxies) and the movements expected 
from the amount of matter observed directly in such systems. This 
problem, also noticed when one analyses the data obtained from the 
cosmic microwave background \cite{WMAP}, was christened as the problem 
of dark matter.

Therefore, today's standard models of cosmology must face the above two 
problems and, therefore, the models used possess in general five basic 
components: photons, baryons, neutrinos, dark matter and dark energy. 
The question ``what are dark matter and dark energy?'' has, in such 
models, a standard answer \cite{DM}: weakly interacting massive 
particles (WIMPs) combined either to a cosmological constant 
($\Lambda$-CDM models) or to some kind of scalar field (quintessence 
models). Both cosmological constant and a quintessential field could be 
the source of energy for the acceleration of the universe, while the 
weakly interacting massive particles, non-luminous, would be responsible 
for the effect of dark matter.

Another kind of proposal consists in supposing that maybe dark matter 
and dark energy are in fact different manifestations of a single exotic 
entity, a component named quartessence (or unified dark matter, UDM) 
\cite{Kamenshchik}-\cite{quart11}. Quartessential models possess a single 
fluid whose main characteristic is to present an exotic equation of 
state $p=p\left(\rho\right)$, which leads to a positive pressure in 
early phases of the universe and to negative pressures in the late 
phases. In such context, the main goal of this article is to present a 
study indicating how density perturbations can evolve in a universe 
composed solely of a single quartessential fluid. Here, the specific 
model used is known as the ``modified Chaplygin gas'' 
\cite{Benaoum}--\cite{mcg2}. It is very important to notice that 
solutions for the scale factor obtained in this model may be obtained 
also in another kinds of cosmological models \cite{EueMartin}. 
Therefore, the study of the evolution of perturbations in each one of 
these models is an important step in the direction of `breaking the 
degeneracy' between them.

The structure of this work is the following: first, the modified 
Chapligyn gas model is presented, followed by another section with a 
brief -- and somewhat pedagogical -- review of the theory concerning the 
evolution of perturbations, leading to a differential equation which is 
used, in sequence, to show how the perturbations evolve in classical 
single fluid models. In the main section of the article the evolution of 
perturbations for a modified Chaplygin gas model \cite{Sandro} is 
obtained, with conclusions being presented in the last section. 
Throughout the text natural units are used, where $c=G=\hbar=k_B=1$, 
unless stated otherwise. Also, arbitrary constants present in 
mathematical solutions are always labeled as $c_1$ and $c_2$.

\section{\label{mCg}The modified Chaplygin gas model}

The modified Chaplygin gas is characterized by the equation of state
\begin{equation}
\label{equationmCg}
p=\left(\gamma-1\right)\rho-M\rho^{-\mu}\,,
\end{equation}
where $\gamma$, $M$ and $\mu$ are free parameters. For $\gamma=\mu=1$ one 
has the original Chaplygin gas, while for $M=0$ one obtains the usual 
linear equation of state. One must notice that such gas is an ideal fluid, in the sense that it does not present variations in space, ie, $\rho=\rho\left( t\right)$. Therefore, the modified Chaplygin gas model does not incorporate any anisotropic stress, what distinguishes it from some more elaborate models \cite{Koivisto}.

Using the above equation, together with the condition for conservation 
of energy in an expanding universe,
\begin{equation}
\dot{\rho}+3\frac{\dot{a}}{a}\left(p+\rho\right)=0\,,
\end{equation}
one obtains the relation
\begin{equation}
\rho=\left[\frac{M}{\gamma}+\left(\rho_0^{1+\mu}-\frac{M}{\gamma}\right)\left(\frac{a}{a_0}\right)^{-3\gamma\left(1+\mu\right)}\right]^{\frac{1}{1+\mu}}\,,
\end{equation}
which, when applied in the Friedmann equation,
\begin{equation}
\left(\frac{\dot{a}}{a}\right)^2+\frac{k}{a^2}=\frac{8\pi}{3}\rho\,,
\end{equation}
yields general analytical solutions, valid for any value of the curvature parameter $k$, only for some values of $\gamma$ and $\mu$. For example, 
for a flat space, Debnath, Banerjee and Chakraborty \cite{DBC} give the 
general formula
\begin{eqnarray}
\,_2F_1\left[x,x;1+x;-\frac{M}{\rho_0^{1+\mu}\gamma-M}\left(\frac{a}{a_0}\right)^{3\gamma\left(1+\mu\right)}\right]=
\left(\frac{a}{a_0}\right)^{-\frac{3\gamma}{2}}t\sqrt{6\pi}\gamma \left(\rho_0^{1+\mu}-\frac{M}{\gamma}\right)^{\frac{1}{2\left(1+\mu\right)}}\,,
\end{eqnarray}
where $x=1/\left[2\left(1+\mu\right)\right]$, which can be inverted 
easily to yield $a=a\left(t\right)$ only for $\mu=0$ and $\mu=-1/2$, 
while for spaces with any curvature, explicit solutions for 
$a\left(t\right)$ for the particular choices $\mu=-1/2$, $\gamma=2/3$ 
and $\gamma=4/3$ are given by Costa \cite{Sandro}.

It is usual to choose $\mu>0$, but here we will restrict the analysis to 
the particular choice $\mu=-1/2$, when one has explicitly, 
\begin{equation}
\label{rho1/2}
p=\left(\gamma -1\right)\rho-M\rho^{1/2}\,,
\end{equation}
what gives, for a flat space, where the curvature is null,
\begin{equation}
\label{rho1/22}
\rho=\left[\frac{M}{\gamma}+\left(\rho_0^{1/2}-\frac{M}{\gamma}\right)\left(\frac{a}{a_0}\right)^{-\frac{3\gamma}{2}}\right]^{2}=\frac{3H^2}{8\pi}\,,
\end{equation}
and, consequently,
\begin{equation}
\label{at}
a\left(t\right)=\lambda\left(e^{\alpha t}-1\right)^{\frac{2}{3\gamma}}\,,
\end{equation}
where $\lambda$ is a constant linked to $\gamma$, $M$, $\rho_0$ and 
$a_0$, and $\alpha=\sqrt{6\pi M^2}$. Therefore, in such model
\begin{equation}
\ddot{a}=\frac{2\lambda}{3\gamma}\alpha^2e^{\alpha t}\left(e^{\alpha t}-1\right)^{\frac{2}{3\gamma}-2}\left(\frac{2}{3\gamma}e^{\alpha t}-1\right)\,,
\end{equation}
implying that if $e^{\alpha t}>3\gamma/2$ one will have $\ddot{a}>0$. 
The conclusion is that values of $\gamma$ greater than $2/3$ are more 
realistic in this model, since then the universe will experience an 
acceleration on its expansion only in later moments.

Observing equations (\ref{rho1/2}) and (\ref{rho1/22}) and considering the final argument of the previous paragraph, a `good' choice could be $\gamma=4/3$, since in such 
case the equation of state would interpolate from a phase similar to a radiation 
dominated universe ($\rho\propto a^{-4}$, for 
$a<<1$) to a 
phase with negative pressure, causing the acceleration of the expansion (for $\rho<9M^2$), presenting also a phase of pressureless matter (since for values of $\rho\approx 9M^2$, one has $p\approx 0$). In resume, then, this particular modified Chaplygin gas model, with 
$\mu=-1/2$ and $\gamma=4/3$, could, at least in principle, play the role 
of an effective equation of state for a mixture of radiation, matter and 
a dynamical cosmological constant.

It is very important to remark that the solution for the scale factor 
given by eq. (\ref{at}) (the background solution) can be obtained 
also in another kinds of flat cosmological models, such as in a decaying vaccum cosmology 
\cite{SauloeAlcaniz}. Since an important difference between these two 
models is on the equation of state used in each one of them, a 
cosmological test based on properties affected directly by the equation 
of state could be a potential way to distinguish the models. The 
analysis of the evolution of perturbations, which has a dependence on 
the equation of state, is, therefore, a possible way to characterize a 
model in a way enough to separate it from its concurrents. In the next 
section, then, we present an overview of the general theory used for 
describing the evolution of perturbations of density, in order to show 
the particular relevance of the equation of state in such approach.

\section{The theory for the evolution of perturbations}

To make a comprehensive overview of the theory concerning the evolution 
of perturbations in an expanding universe, it is safe to begin with a 
Newtonian approach, where the radial acceleration $\ddot{R}$ suffered by 
a test particle previously in equilibrium on the surface of a 
homogeneous sphere of radius $R$, due to the gravity caused by the 
increase in the sphere of an excess of mass $\Delta m$, is given by the 
expression \cite{Ryden}
\begin{equation}
\label{ddotR}
\ddot{R}=-\frac{G}{R^2}\Delta m=-\frac{G}{R^2}\left(\frac{4\pi}{3}R^3\bar{\rho}\delta\right)\,,
\end{equation}
where $\bar{\rho}$ is the mean density of the sphere and $\delta$ is a 
density fluctuation (known also as density contrast), such that the 
density inside the sphere is given as 
$\rho=\bar{\rho}\left(1+\delta\right)$, with $\delta<<1$. The total mass 
of the sphere is $m=4\pi\bar{\rho}\left(1+\delta\right)/3$, from where 
one can write, approximately,
\begin{equation}
\label{Rt}
R\left(t\right)\approx\left(\frac{3m}{4\pi\bar{\rho}}\right)^{1/3}\left[1-\frac{1}{3}\delta\left(t\right)\right]=R_0\left[1-\frac{1}{3}\delta\left(t\right)\right]\,.
\end{equation}

Using the relations given by equations (\ref{ddotR}) and (\ref{Rt}), and 
noticing that $R\approx R_0$, one can write, then,
\begin{equation}
\frac{\ddot{R}}{R}\approx -\frac{\ddot{\delta}}{3}\,,
\end{equation}
and, therefore,
\begin{equation}
\label{Newton}
\ddot{\delta}= 4\pi G\bar{\rho}\delta\,.
\end{equation}
This is the equation governing the evolution of density perturbations in 
a Newtonian static universe. However, if the universe is expanding, one 
can generalize the above arguments to obtain an equation where the scale 
factor appears explicitly \cite{Ryden},
\begin{equation}
\label{expansion}
\ddot{\delta}+2\frac{\dot{a}}{a}\dot{\delta}=4\pi G\bar{\rho}\delta\,.
\end{equation}

The procedure leading to a fully general relativistic equation having 
the density contrast $\delta$ as variable is a more elaborate one. Its 
development can be traced back to Hawking \cite{Hawking}, with further 
advances by Olson \cite{Olson} and, more recently, by Lyth and Mukherjee 
\cite{Lyth1}, and Lyth and Stewart \cite{Lyth2}. A resume of it can be 
found in a textbook by Padmanabhan \cite{Padmanabhan}. Here we just cite 
the final result, which is, explicitly, for a 
Friedamnn-Lema\^{i}tre-Robertson-Walker (FLRW) flat 
universe\footnote{See Appendix {\ref{ap1}} for another forms of this 
result.} (where the curvature parameter is zero),
\begin{eqnarray}
\label{Pad1}
\ddot{\delta}&+&H\left[2-3\left(2w-v^2\right)\right]\dot{\delta}
-\frac{3}{2}H^2\left(1-6v^2-3w^2+8w\right)\delta=-k^2\frac{v^2}{a^2}\delta\,,
\end{eqnarray}
where $H=\dot{a}/a$ is the Hubble parameter, $w=p/\rho$ is the ratio 
between pressure and energy density of the cosmological fluid, 
$v^2=\frac{\partial p}{\partial\rho}$ is the square of the sound 
velocity in the fluid, and $k$ is the wavenumber of the Fourier mode of 
the perturbation. Notice that now one has a dependence on the equation 
of state $p=p\left(\rho\right)$ of the cosmological fluid in study, 
included implicitly in the definitions of $\omega$ and $v^2$.

To better understand the processes described by these differential 
equations one can use a purely pragmatic point of view to see whether 
they can be obtained from effective ``Lagrangian densities'' where the 
degrees of freedom are given by $\delta$ and $\dot{\delta}$. From the 
Lagrangian density
\begin{equation}
\label{lag1}
\mathcal{L}_0=\frac{c^2}{G}\left(\frac{\dot{\delta}^2}{2}+2\pi G\bar{\rho}\delta^2\right)=\frac{1}{2}\left(\frac{c^2\dot{\delta}^2}{G}\right)+2\pi\bar{\rho}c^2\delta^2
\end{equation}
one produces the Newtonian equation (\ref{Newton}), with the constants 
$c$ and $G$ being used to give the correct dimensionality. In this 
Lagrangian one has a usual kinetic term, proportional to 
$\dot{\delta}^2$, and a {\it negative} potential proportional to 
$\delta^2$, which implies there are no stable points of equilibrium. The 
solutions, thus, represent only decaying or growing modes.

The result for an expanding universe, eq. (\ref{expansion}), can also be 
obtained by a Lagrangian density\footnote{One way to build this result 
is to {\it assume} that the Lagrangian density given in equation 
(\ref{lag1}) is an invariant quantity. Since the fraction $c^2/G$ is 
invariant, and distances grow in an expanding universe, while 
frequencies are shortened, the remaining quantity, which has dimensions 
of a squared frequency, must be multiplied by the square of the scale 
factor.}
\begin{equation}
\label{lag2}
\mathcal{L}=\mathcal{L}_0a^2=\frac{c^2}{G}\left(\frac{a^2\dot{\delta}^2}{2}+\frac{3a^2H^2\delta^2}{4}\right)\,,
\end{equation}
with $H^2$ standing in the place of $8\pi G\bar{\rho}/3$. Again the 
potential is negative, but since one has now the kinetic term 
proportional to $a^2\dot{\delta}^2$, there is a `friction' in the 
system.

It is not hard to verify that the general relativistic equation 
(\ref{Pad1}), valid for a flat space, can be obtained from a 
generalization of the above Lagrangians,
\begin{eqnarray}
\label{lag3}
\mathcal{L}=\frac{\hbar}{c^3}\frac{\sqrt{-g}\mathcal{R}}{1+\omega}\frac{a^2}{2}
\times\left[\dot{\delta}^2+\frac{3H^2\delta^2}{2}\left(1+8\omega-3\omega^2-6v^2-\frac{2v^2k^2}{3a^2H^2}\right)\right]\,,
\end{eqnarray}
where $g$ is the determinant of the metric, and 
$\mathcal{R}=K^2-K_{ij}K^{ij}-\,^3\mathcal{R}$ is the Ricci scalar built 
from the extrinsic curvature tensor $K_{ij}$ and the three-curvature 
$\,^3\mathcal{R}$ \cite{Kolb}. For a flat FLRW model $g=-a^6$ and 
$\mathcal{R}=6H^2$. Again the potential is proportional to $\delta^2$ 
but it can be positive or negative, according to the value of a 
polynomial expression involving $\omega$ and $v^2$. Therefore, in the 
general relativistic case one cannot say that all solutions have only 
growing and decaying modes, since one can have oscillatory behavior. In 
order to make more clear this point, we present some examples in the 
next subsection.

\subsection{Perturbations in classical models of a single fluid}

As commented before, the evolution of perturbations in the general 
relativistic case can present three distinct possible behaviours: 
growing, decaying and oscillatory. This can be seen in the simplest 
models of cosmology, which use a linear equation of state,
\begin{equation}
p=\left(\gamma-1\right)\rho\,,
\end{equation}
where $\gamma\neq 0$ is a numerical factor. Such relation, when applied 
into the energy conservation equation,
\begin{equation}
\dot{\rho}+3H\left(p+\rho\right)=0\,,
\end{equation} 
and in conjunction with the Friedmann equation for the flat space,
\begin{equation}
H^2=\frac{8\pi}{3}\rho\,,
\end{equation}
yields, for the scale factor,
\begin{equation}
a\left(t\right)=a_0\left(\frac{t}{t_0}\right)^{\frac{2}{3\gamma}}\,,
\end{equation}
and, consequently,
\begin{equation}
H=\frac{2}{3\gamma}\frac{1}{t}\,.
\end{equation} 

Using these results into equation (\ref{Pad1}), one then ends with the 
equation
\begin{eqnarray}
\ddot{\delta}+\frac{2}{3\gamma t}\left(5-3\gamma\right)\dot{\delta}&+&\frac{2}{3\gamma^2t^2}\left(3\gamma^2-8\gamma +4\right)\delta=
-k^2\frac{t_0^{\frac{4}{3\gamma}}}{a_0^2t^{\frac{4}{3\gamma}}}\left(\gamma -1\right)\delta\,.
\end{eqnarray}
For any value of $\gamma\neq 0$ this equation has analytical solutions, 
since it is a particular case of equation {\bf 8.491.12} from Gradshteyn 
and Ryzhik \cite{Gradshteyn},
\begin{equation}
z^2u^{\prime\prime}+\left(2\alpha-2\beta\nu+1\right)zu^{\prime}+\left[\beta^2\mu^2z^{2\beta}+\alpha\left(\alpha-2\beta\nu\right)\right]u=0\,,
\end{equation} 
with $$\beta=1-\frac{2}{3\gamma}\,, 
\,\,\mu^2=\frac{k^2t_0^{\frac{4}{3\gamma}}}{a_0^2}\left(\gamma-1\right)\left(1-\frac{2}{3\gamma}\right)^{-2}\,,$$ 
$$\alpha=-1+\frac{2}{\gamma}\,\,\,\mathrm{and}\,\,\, 
\nu=\left(\frac12+\frac{1}{3\gamma}\right)\left(1-\frac{2}{3\gamma}\right)^{-1}$$ 
{or} $$\alpha=-2+\frac{4}{3\gamma}\,\,\,\mathrm{and}\,\,\, 
\nu=-\left(\frac12+\frac{1}{3\gamma}\right)\left(1-\frac{2}{3\gamma}\right)^{-1}\,.$$

\subsubsection{Dust matter: $\gamma=1$}

In this case the equation to be solved is
\begin{equation}
\ddot{\delta}+\frac{4}{3t}\dot{\delta}-\frac{2}{3t^2}\delta=0\,,
\end{equation}
with the solution
\begin{equation}
\delta=c_1t^{-1}+c_2t^{2/3}\,,
\end{equation}
where the $c_i$ are arbitrary constants. Here, therefore, the solution 
obtained contains only a growing and a decreasing mode.

\subsubsection{Radiation: $\gamma=4/3$}

In this case, one has
\begin{equation}
\label{radiation}
\ddot{\delta}+\frac{1}{2t}\dot{\delta}-\frac{1}{2t^2}\delta=-\frac{k^2}{3}\frac{t_0}{a_0^2t}\delta\,,
\end{equation}
with the {\it oscillatory} solution
\begin{eqnarray}
\delta=c_1\left(\frac{\sin b_0t^{1/2}}{b_0t^{1/2}}-\cos b_0t^{1/2}\right)
+c_2\left(\frac{\cos b_0t^{1/2}}{b_0t^{1/2}}+\sin b_0t^{1/2}\right)\,,
\end{eqnarray}
where $b_0^2=4k^2t_0\left(3a_0^2\right)^{-1}$ and, again, the $c_i$ are 
arbitrary constants.

\section{Evolution of perturbations in the modified Chaplygin gas model}

As seen in Section \ref{mCg}, the model we will study is defined by 
the equation of state (\ref{rho1/2}),
$$
p=\left(\gamma-1\right)\rho-M\rho^{1/2}\,,
$$
which has two free parameters, $\gamma$ and $M$. In order to 
verify how the perturbations in the matter density evolve in such model 
we need to calculate the quantities $\omega=p/\rho$ and 
$v^2=\frac{\partial p}{\partial \rho}$ which appear in equation 
(\ref{Pad1}). Therefore, in the model analysed here one has explicitly
\begin{equation}
w=\frac{p}{\rho}=\gamma-1-M\rho^{-1/2}=\gamma-1-\frac{2\alpha}{3H}\,,
\end{equation}
\begin{equation}
v^2=\frac{\partial p}{\partial \rho}=\gamma-1-\frac{M}{2}\rho^{-1/2}=\gamma-1-\frac{\alpha}{3H}\,.
\end{equation}
Substituting the results above in the equation for perturbations, one 
has
\begin{eqnarray}
\label{principal0}
\ddot{\delta}+\left[\left(5-3\gamma\right)H+3\alpha\right]\dot{\delta}+
\left[\left(3\gamma^2-8\gamma+4\right)H^2
+\left(\frac{22}{3}-4\gamma\right)\alpha H+\frac{4}{3}\alpha^2\right]\frac{3\delta}{2}=\nonumber\\
-\frac{k^2}{\lambda^2}\left(\frac{3\gamma}{2\alpha}H-1\right)^{\frac{4}{3\gamma}}\left(\gamma-1-\frac{\alpha}{3H}\right)\delta\,.
\end{eqnarray}
It is easy to notice that this equation simplifies for $\gamma=4/3$. 
Then, this one will be the choice used in the next subsection. Anyway, 
one can rewrite the above equation using a new variable, $y=1-\frac{3\gamma 
H}{2\alpha}=\left(1-e^{\alpha t}\right)^{-1}$, to produce
\begin{eqnarray}
\label{principaly}
y\left(1-y\right)\frac{d^2\delta}{dy^2}+\left[\left(\frac{10}{3\gamma}-4\right)y-\frac{10}{3\gamma}\right]
\frac{d\delta}{dy}
+\left(\frac{2}{1-y}-\frac{6\gamma^2-16\gamma+8}{3\gamma^2}+\frac{8+6\gamma}{3\gamma^2y}\right)\delta=\nonumber\\
-\frac{k^2}{\lambda^2\alpha^2}\left(-y\right)^{\frac{4}{3\gamma}}\left[\left(1-\frac{\gamma}{2}\right)\left(\frac{1}{1-y}+\frac{1}{y}\right)+\frac{\gamma}{2}\frac{1}{\left(1-y\right)^2}\right]\delta\,.
\end{eqnarray}
For the particular case of the mode with $k=0$, this last equation has a general solution for any value of $\gamma\neq 
0$,
\begin{equation}
\label{eqz}
\delta\left(y\right)=c_1\frac{y^{1+\frac{4}{3\gamma}}}{1-y}\,_2F_1\left(2,1-\frac{2}{3\gamma};
2-\frac{2}{3\gamma};y\right)+c_2\frac{y^{\frac{2}{\gamma}}}{1-y}\,,
\end{equation}
where the $c_i$s are arbitrary constants.

\subsection{Modified Chaplygin gas model with $\gamma=4/3$}

For the particular choice $\gamma=4/3$, equation (\ref{principal0}) 
becomes
\begin{eqnarray}
\label{principal}
\ddot{\delta}+\left(H+3\alpha\right)\dot{\delta}-\left(H-2\alpha\right)\left(2H+\alpha\right)\delta=
-\frac{k^2}{3\lambda^2}\left(1-\frac{\alpha}{H}\right)\left(\frac{2H}{\alpha}-1\right)\delta\,.
\end{eqnarray}
In the particular case of the mode $k=0$ one has a solution given by eq. (\ref{eqz}),
\begin{eqnarray}
\label{solk0}
\delta\left(t\right)=c_1\frac{1-e^{-\alpha t}}{\left(1-e^{\alpha
t}\right)^{2}}\,_2F_1\left(2,\frac{1}{2};\frac{3}{2};\frac{1}{1-e^{\alpha
t}}\right)
+c_2\frac{1-e^{-\alpha t}}{\left(1-e^{\alpha
t}\right)^{\frac{3}{2}}}\,.
\end{eqnarray}
This solution, which {\it is not} oscillatory, can also be written in 
terms of the Hubble parameter $H$, as
\begin{equation}
\delta\left(H\right)=c_1\delta_1+c_2\delta_2\,,
\end{equation}
where
\begin{equation}          
\delta_1=\frac{\left(2H-\alpha\right)^{\frac32}}{H}
\end{equation}
and
\begin{equation}
\delta_2=\frac{\left(2H-\alpha\right)^2}{2\alpha H^2}+
\frac{1}{H}\left(\frac{2H}{\alpha}-1\right)^{\frac32}\arctan\left(\frac{2H}{\alpha}-1\right)^{\frac12}\,,
\end{equation}
and again the $c_i$ are arbitrary constants.

\subsubsection{Approximate solutions}

An approximate solution valid for small values of $t$ can be obtained 
directly from eq. (\ref{principal}), using that $a\left(t\right)\approx 
\lambda\alpha^{1/2} t^{1/2}$ and $H\approx t^{-1/2}$, so that the 
equation to be solved becomes
\begin{eqnarray}
\ddot{\delta}+\left(1+6\alpha t\right)\frac{\dot{\delta}}{2t}+\left(4\alpha^2 t^2+3\alpha t-1\right)\frac{\delta}{2t^2}=
-\frac{k^2}{3\lambda^2\alpha t}\left(1-3\alpha t+2\alpha^2t^2\right)\delta\,,
\end{eqnarray}
or, approximately,
\begin{equation}
\ddot{\delta}+\frac{\dot{\delta}}{2t}-\frac{\delta}{2t^2}=-\frac{k^2}{3\lambda^2\alpha t}\delta\,,
\end{equation}
which is the equation for a classical model with $\gamma=4/3$, eq. 
(\ref{radiation}), with $\lambda^2\alpha$ taking the place of 
$a_0^2t_0^{-1}$.

In order to obtain an approximate solution valid for large values of $t$ 
it is interesting to rewrite equation (\ref{principal}) as
\begin{eqnarray}
\ddot{\delta}&+&\frac{\alpha}{2}\left(7+\frac{1}{e^{\alpha t}-1}\right)\dot{\delta}+\frac{\alpha^2}{2}\left[6+\frac{e^{\alpha t}-2}{\left(e^{\alpha t}-1\right)^2}\right]\delta =
-\frac{k^2}{3\lambda^2}e^{-\alpha t}\left(\frac{1}{e^{\alpha t}-1}-1\right)\delta\,.
\end{eqnarray}
Now, using that $\delta\left(t\right)=e^{-\alpha t}\left(e^{\alpha 
t}-1\right)^{-1/2} \Gamma\left(t\right)$, the remaining equation is
\begin{equation}
\left(e^{\alpha t}-1\right)\ddot{\Gamma}+\frac{\alpha}{2}\left(e^{\alpha t}-2\right)\dot{\Gamma}=-\frac{k^2}{3\lambda^2}e^{-\alpha t}\left(2-e^{\alpha t}\right)\Gamma \,,
\end{equation}
which, for large values of $\alpha t$, can be approximed to
\begin{equation}
e^{\alpha t}\left(\ddot{\Gamma}+\frac{\alpha}{2}\dot{\Gamma}\right)=\frac{k^2}{3\lambda^2}\Gamma \,,
\end{equation}
with the exact solutions
\begin{equation}
\Gamma\left(t\right)=c_1\exp\left[{\frac{2\sqrt{3}}{3}\frac{k}{\lambda\alpha}e^{-\frac{\alpha t}{2}}}\right]+c_2\exp\left[{-\frac{2\sqrt{3}}{3}\frac{k}{\lambda\alpha}e^{-\frac{\alpha t}{2}}}\right]\,,
\end{equation}
valid for $k\neq 0$, and
\begin{equation}
\Gamma\left(t\right)=c_1-c_2\frac{2}{\alpha}e^{-\frac{\alpha t}{2}}\,,
\end{equation}
valid for $k=0$, where the $c_i$ are constants.

Then, summarizing the approximate results for the modes with $k\neq 0$, 
one has, for $t<<1$,
\begin{eqnarray}
\label{tmenor}
\delta_{t<<1}=c_1\left(\frac{\sin bt^{1/2}}{bt^{1/2}}-\cos bt^{1/2}\right)
+c_2\left(\frac{\cos bt^{1/2}}{bt^{1/2}}+\sin bt^{1/2}\right)\,,
\end{eqnarray}
where now $b^2=4k^2\left(3\lambda^2\alpha\right)^{-1}$, while, in the 
other extreme, for $e^{\alpha t}>>1$,
\begin{eqnarray}
\delta_{t>>1}=c_1\exp\left({\frac{2\sqrt{3}}{3}\frac{k}{\lambda\alpha}e^{-\frac{\alpha t}{2}}}-\frac{3\alpha t}{2}\right)
+c_2\exp\left({-\frac{2\sqrt{3}}{3}\frac{k}{\lambda\alpha}e^{-\frac{\alpha t}{2}}}-\frac{3\alpha t}{2}\right),
\end{eqnarray}
with the $c_i$ being arbitrary constants.

\subsubsection{Solutions in terms of scale factor and conformal time}

Although it is more interesting to see the density contrast as a 
function of the cosmological time $t$, it may be also useful to study 
its evolution in terms of other variables, such as the conformal time 
$\eta$ or the scale factor $a$. The Appendix \ref{ap1}, at the end of 
this work, shows such analysis for the classical single fluid models. 
Since in those models there is a substantial simplification of the 
differential equations for the density contrast, one can see whether or 
not this is also the case in the model of a modified Chaplygin gas.

We choose to begin this extra analysis using the scale factor as the 
independent variable, when the relevant variable transformation is
\begin{equation}
dt=\frac{2a}{\alpha\left(a^2+\lambda^2\right)}da\,,
\end{equation}
and, therefore, the equation to be solved becomes
\begin{eqnarray}
\label{principal2a}
\left(a^2+\lambda^2\right)\frac{d^2\delta}{da^2}+8a\frac{d\delta}{da}-\frac{4k^2}{3\alpha^2}\frac{a^2-\lambda^2}{\left(a^2+\lambda^2\right)^2}\delta=-2\left(1-\frac{\lambda^2}{a^2}+\frac{6a^2+\lambda^2-a^2}{a^2+\lambda^2}\right)\delta\,.
\end{eqnarray}
Using the hypothesis
\begin{equation}
\delta=\frac{\lambda^2}{a\left(a^2+\lambda^2\right)}\Gamma
\end{equation}
one gets
\begin{equation}
\label{principal3}
\left(a^2+\lambda^2\right)\frac{d^2\Gamma}{da^2}+\frac2a\left(a^2-\lambda^2\right)\frac{d\Gamma}{da}=\frac{4k^2}{3\alpha^2}\frac{a^2-\lambda^2}{\left(a^2+\lambda^2\right)^2}\Gamma\,,
\end{equation}
which for $a\rightarrow 0$ yields
\begin{equation}
\frac{d^2\Gamma}{da^2}-\frac2a\frac{d\Gamma}{da}+\frac{4k^2}{3\alpha^2\lambda^4}\Gamma=0\,,
\end{equation}
with the solution
\begin{equation}
\Gamma_{a\rightarrow 0}=c_1\left(\frac{\sin b_1 a}{b_1}-a\cos b_1 a\right)+c_2\left(\frac{\cos b_1 a}{b_1}+a\sin b_1 a\right)\,,
\end{equation}
where $b_1^2=4k^2/(3\alpha^2\lambda^4)$, while for $a>>\lambda$ one has
\begin{equation}
\frac{d^2\Gamma}{da^2}+\frac2a\frac{d\Gamma}{da}-\frac{4k^2}{3\alpha^2a^4}\Gamma=0\,,
\end{equation}
with the solution
\begin{equation}
\Gamma_{a>>\lambda}=c_1e^{b_1\lambda^2/a}+c_2e^{-b_1\lambda^2/a}\,.
\end{equation}
Finally, for $k=0$ there is an exact solution,
\begin{equation}
\label{Ak0_1}
\Gamma_{k=0}=c_1\left[\frac{1}{2\lambda}\arctan\left(\frac{a}{\lambda}\right)-\frac{a}{2\left(a^2+\lambda^2\right)}\right]+c_2\,.
\end{equation}
Therefore, in resume, one has
\begin{eqnarray}
\delta_{a\rightarrow 
0}=\frac{c_1}{\left(a^2+\lambda^2\right)}\left(\frac{\sin b_1 a}{b_1 a}-\cos b_1 a\right)
+\frac{c_2}{\left(a^2+\lambda^2\right)}\left(\frac{\cos b_1 a}{b_1 a}+\sin b_1 a\right)\,,
\end{eqnarray}
\begin{equation}
\delta_{a>>\lambda}=\frac{\lambda^2}{a\left(a^2+\lambda^2\right)}\left[c_1e^{b_1 \lambda^2/a}+c_2e^{-b_1 \lambda^2/a}\right]\,,
\end{equation}
and
\begin{eqnarray}
\delta_{k=0}=c_1\left[\frac{\lambda\arctan\left(\frac{a}{\lambda}\right)}{2a\left(a^2+\lambda^2\right)}-\frac{\lambda^2}{2\left(a^2+\lambda^2\right)^2}\right]
+c_2\frac{\lambda^2}{a\left(a^2+\lambda^2\right)}\,.
\end{eqnarray}

We can also present equations for evolution of the perturbations in 
terms of the conformal time,
\begin{equation}
\eta=\frac{2}{\lambda\alpha}\arccos\left(e^{-\frac{\alpha t}{2}}\right)\,,
\end{equation}  
so that 
\begin{equation}
a=\lambda\tan\left(\frac{\lambda\alpha\eta}{2}\right)\,,
\end{equation}
with $0\leq\eta<\frac{\pi}{2}$. With such substitutions equation 
(\ref{principal3}), for example, becomes
\begin{equation}
\label{principal4}
\frac{d^2\Gamma}{d\eta^2}-\lambda\alpha\cot\left(\frac{\lambda\alpha\eta}{2}\right)\frac{d\Gamma}{d\eta}=\frac{k^2}{3}\left[1-2\cos^2\left(\frac{\lambda\alpha\eta}{2}\right)\right]\Gamma\,,
\end{equation}
with the solution
\begin{equation}
\Gamma_{k=0}=\frac{c_1}{2}\left[\eta-\frac{\sin\left(\lambda\alpha\eta\right)}{\lambda\alpha}\right]+c_2\,,
\end{equation}
for $k=0$, which is a result completely equivalent to equation 
(\ref{Ak0_1}).

Equation (\ref{principal4}) produces a final important result, which can 
be obtained changing the coordinate to $z=e^{-\frac{\alpha 
t}{2}}=\cos\left(\frac{\lambda\alpha\eta}{2}\right)$. Doing this, the 
equation becomes
\begin{equation}
\left(1-z^2\right)\frac{d^2\Gamma}{dz^2}+z\frac{d\Gamma}{dz}=\frac{4k^2}{3\lambda^2\alpha^2}\left(1-2z^2\right)\Gamma,
\end{equation}
which is a {\it spheroidal wave equation} of the type
\begin{equation}
\left(1-z^2\right)\frac{d^2w}{dz^2}-2\left(a+1\right)z\frac{dw}{dz}+\left(b-c^2z^2\right)w=0\,,
\end{equation}
as defined by Stratton \cite{Stratton1,Stratton2}, with $a=-3/2$, 
$b=-\frac{4k^2}{3\lambda^2\alpha^2}$ and $c^2=-2b$. Solutions to this 
equation are usually given in terms of elaborate series of special 
functions or orthogonal polynomials, such as \cite{Morse,Leaver}
\begin{equation}
w\left(z\right)=\sum_{n^\prime} d_{n}T_n^a\left(z\right)\,,
\label{series}
\end{equation} 
where the $d_n$'s are the cofficients of the expansion, and 
$T^{m}_{n}\left(z\right)$ are the Gegenbauer polynomials related to the 
hypergeometric series $\,_2F_1$ by
\begin{equation}
T^{m}_{n}\left(z\right)=\frac{\left(n+2m\right)!}{2^mn!m!}\,_2F_1\left(n+2m+1,-n;m+1;\frac{1-z}{2}\right)\,.
\end{equation}
The series (\ref{series}) is convergent only for a certain discrete set 
of values of the quantity $A\equiv b+a+a^2$, values which can be 
obtained from a transcendental equation \cite{Abramowitz}, and which can 
be ordered so that for a given $a$, the lowest value of $A$ is labeled 
$A_{a,a}$, the next is $A_{a,a+1}$, generating an index $\ell$ 
associated with the solutions through the eigenvalues 
$A_{a,\ell}<A_{a,\ell+1}$, and so on. The prime on the summation 
indicates inclusion of only even $n$'s when the quantity $\ell-a$ is 
even and only odd $n$'s when $\ell-a$ is odd. Various normalization 
schemes can be used for the $d_n$'s, which satisfy a three-order 
recurrence relation \cite{Mathworld}.

The subject of spheroidal wave functions is a particularly complex one, 
and therefore we will, in sequence, resort to numerical methods in order 
to better visualize the general behaviour of the density perturbations.

\subsubsection{Numerical solutions}

Besides the analytical solutions presented in this article, we also 
performed a numerical integration of Eq. (\ref{principal}) in order to 
analyze the general behavior of the perturbations. The method used for 
the numerical calculation of the second order differential equation was 
the finite difference method. The code used for the numerical 
calculations was tested with the analytical solution obtained with 
$k=0$, say Eq. (\ref{solk0}).

\begin{figure}[ht] 
\begin{center}
\includegraphics*[scale=0.4]{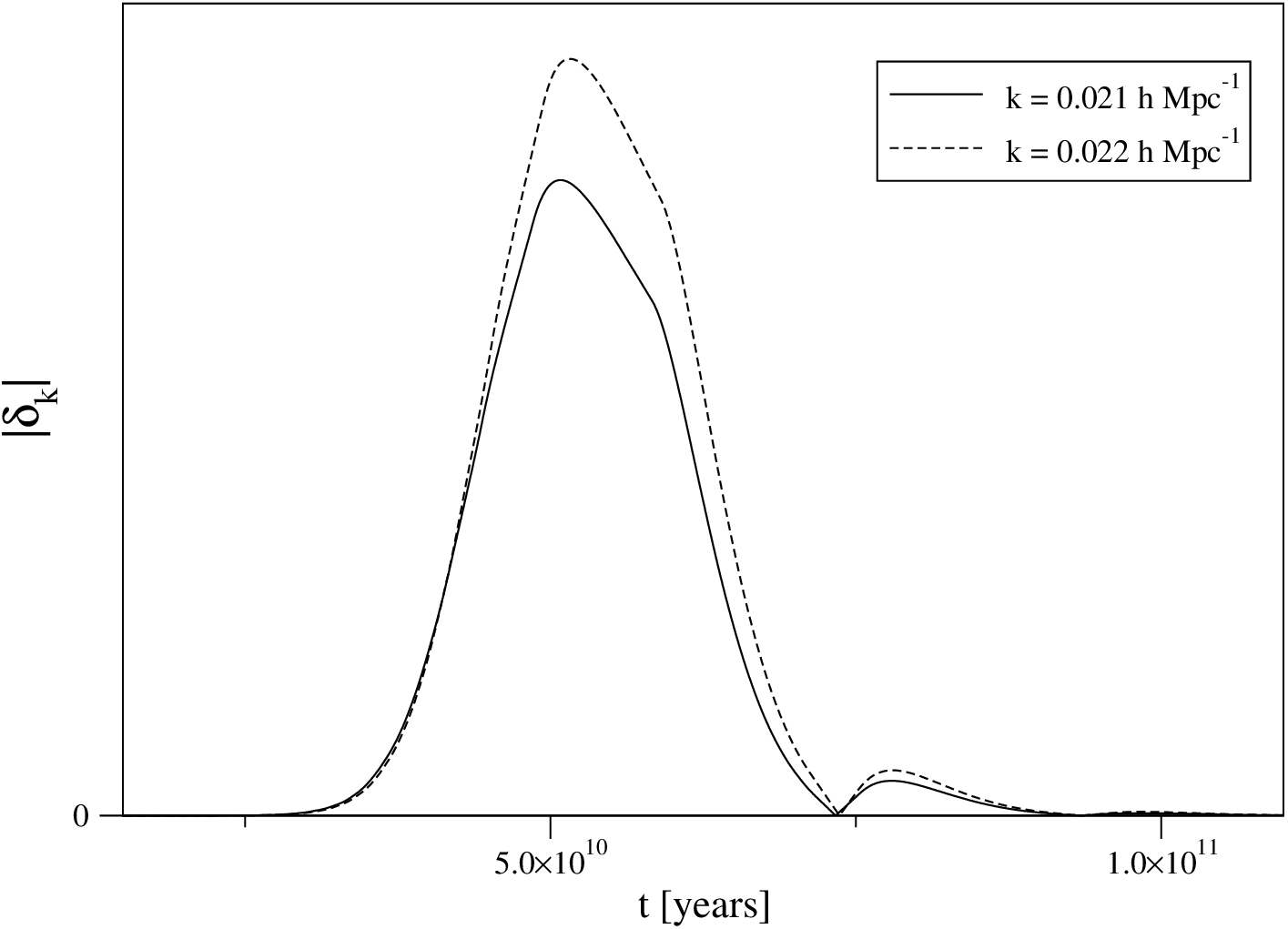}
\end{center}
\caption{Graph showing the general behaviour of perturbations in a 
modified Chapligyn gas with $\gamma=4/3$ for two values of the 
wavenumber $k$. Here, it is clear the presence of peaks. The scale in the vertical axis is arbitrary. Numerical 
results for other wavenumber values present similar behaviours.}
\label{generalbehaviour}
\end{figure}

\begin{figure}[ht]
\begin{center}
\includegraphics*[scale=0.4]{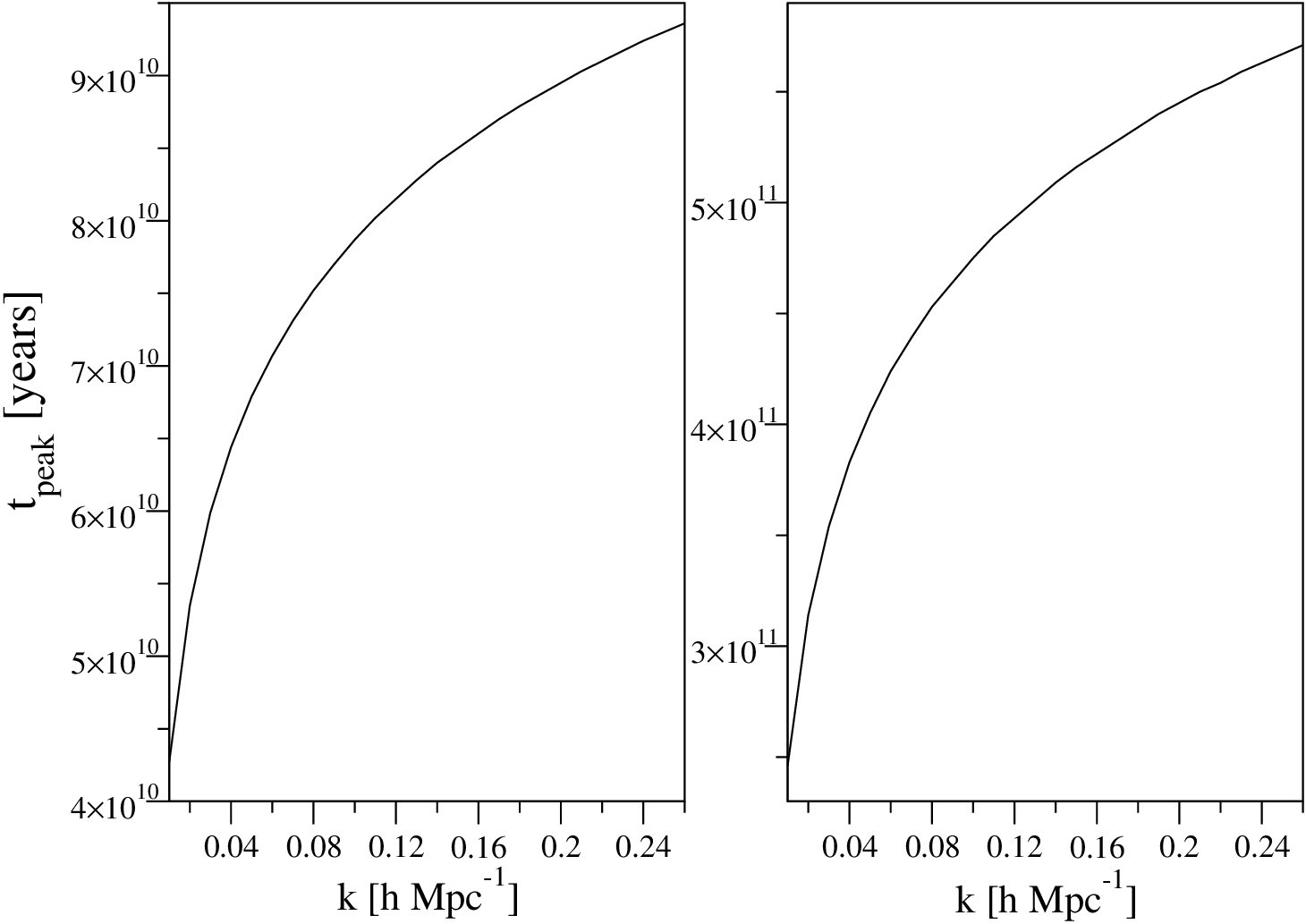}
\end{center}
\caption{Graphs presenting the dependence of the time in which the 
perturbation is maximal, $t_{peak}$, with the wavenumber of the 
perturbation, $k$, for $\gamma=4/3$ (left) and $\gamma=2/3$ (right).}
\label{tpeak}
\end{figure}

We perform the integration in time of Eq. (\ref{principal}) starting 
from $t=10^2$~years to $t=1.2 \times 10^{11}$~years. The approximate 
solution (\ref{tmenor}) was used to apply the initial conditions of our 
problem. The numerical results show clearly the presence of a peak in 
the evolution of the perturbations for any value of $k$. Results for 
some values of the wavenumber $k$ are depicted in Fig.
\ref{generalbehaviour}. The integral path used in Fig. 
\ref{generalbehaviour} was $\Delta t=1$~year. Numerical simulations with 
less acuracy, $\Delta t > 1$~year, show that the time in which the peak 
is maximum is practically the same but some details on the perturbation 
profiles are lost. Note that the maximum value of the perturbation 
increases when we increase the wavenumber value. The time in which the 
perturbation value is maximum also depends on the value of $k$. This 
behaviour is depicted in Fig. \ref{tpeak}. The numerical results of Fig. 
\ref{tpeak} were obtained with a integration path of $\Delta t= 
10^3$~years to save integration time. All figures were built with the 
values $\alpha=1.28\times10^{-10}$ years$^{-1}$, $t_0=1.4\times10^{10}$ 
years and $a_0=a\left(t_0\right)=1$.

\subsection{Modified Chaplygin gas with $\gamma\neq 4/3$}
Another values of $\gamma$ different of $4/3$ allow one to obtain some analytical and numerical results, using procedures analogous to the one used for $\gamma=4/3$. Starting from equation (\ref{principal0}) or (\ref{principaly}), using as variable the scale factor $a$, and putting 
\begin{equation}
\delta=\frac{\left(-1\right)^{\frac{2}{\gamma}}\left(\frac{\lambda}{a}\right)^3}{1+\left(\frac{\lambda}{a}\right)^{\frac{3\gamma}{2}}}\Gamma\,,
\end{equation} 
one obtains
\begin{eqnarray}
\left[1+\left(\frac{a}{\lambda}\right)^{\frac{3\gamma}{2}}\right]\frac{d^2\Gamma}{da^2}-\frac{3\gamma}{2a}\left[1-\left(\frac{a}{\lambda}\right)^{\frac{3\gamma}{2}}\right]\frac{d\Gamma}{da}=-\frac{9\gamma^2k^2}{4\alpha^2a^4}\left\{\frac{\gamma}{2}\frac{\left(\frac{a}{\lambda}\right)^{3\gamma}}{\left[1+\left(\frac{a}{\lambda}\right)^{\frac{3\gamma}{2}}\right]^2}-\left(1-\frac{\gamma}{2}\right)\frac{\left(\frac{a}{\lambda}\right)^{3\gamma}}{1+\left(\frac{a}{\lambda}\right)^{\frac{3\gamma}{2}}}\right\}\Gamma\,.\nonumber\\
\end{eqnarray}

As a first example, for $\gamma=2/3$ one has then
\begin{equation}
\left(a+\lambda\right)\frac{d^2\Gamma}{da^2}+\left(a-\lambda\right)\frac{1}{a}\frac{d\Gamma}{da}=-\frac{k^2\lambda}{3\alpha^2 a^2}\left[\frac{1}{\left(a+\lambda\right)^2}-\frac{2}{\lambda\left(a+\lambda\right)}\right]\Gamma\,,
\end{equation}
which for $a\rightarrow 0$ reduces to
\begin{equation}
\frac{d^2\Gamma}{da^2}-\frac{1}{a}\frac{d\Gamma}{da}-\frac{k^2}{3\alpha^2\lambda^2}\frac{\Gamma}{a^2}=0\,,
\end{equation} 
with the solution
\begin{equation}
\Gamma_{a\rightarrow 0}=a\left[c_1a^{\sqrt{1+\frac{k^2}{3\alpha^2\lambda^2}}}+c_2a^{-\sqrt{1+\frac{k^2}{3\alpha^2\lambda^2}}}\right]\,,
\end{equation}
while for $a>>\lambda$ one has
\begin{equation}
\frac{d^2\Gamma}{da^2}+\frac{1}{a}\frac{d\Gamma}{da}-\frac{2k^2}{3\alpha^2}\frac{\Gamma}{a^4}=0\,,
\end{equation} 
with the solution
\begin{equation}
\Gamma_{a>>\lambda}=c_1K_{0}\left(\sqrt{\frac{2k^2}{3\alpha^2}}\frac{1}{a}\right)+c_2I_{0}\left(\sqrt{\frac{2k^2}{3\alpha^2}}\frac{1}{a}\right)\,.
\end{equation}

As another example, one can use $\gamma=2$ to obtain the equation
\begin{equation}
\left(a^3+\lambda^3\right)\frac{d^2\Gamma}{da^2}+\frac{3}{a}\left(a^3-\lambda^3\right)\frac{d\Gamma}{da}=-\frac{9k^2\lambda^3}{\alpha^2}\frac{a^2}{\left(a^3+\lambda^3\right)^2}\Gamma\,.
\end{equation}
When $a\rightarrow 0$ such equation becomes
\begin{equation}
\frac{d^2\Gamma}{da^2}-\frac{3}{a}\frac{d\Gamma}{da}+\frac{9k^2}{\alpha^2\lambda^6}a^2\Gamma=0\,,
\end{equation}
which has the solution
\begin{equation}
\Gamma_{a\rightarrow 0}=a^2\left[c_1J_1\left(\frac{3k}{2\alpha\lambda^3}a^{2}\right)+c_2J_{-1}\left(\frac{3k}{2\alpha\lambda^3}a^{2}\right)\right]\,,
\end{equation}
while for $a>>\lambda$ one has
\begin{equation}
\frac{d^2\Gamma}{da^2}+\frac{3}{a}\frac{d\Gamma}{da}+\frac{9k^2\lambda^3}{\alpha^2}\frac{\Gamma}{a^7}=0\,,
\end{equation}
which has the solution
\begin{equation}
\Gamma_{a>>\lambda}=a^{-1}\left[c_1J_{\frac{2}{5}}\left(\frac{6k\lambda^{\frac32}}{5\alpha}a^{-\frac{5}{2}}\right)+c_2J_{-\frac{2}{5}}\left(\frac{6k\lambda^{\frac32}}{5\alpha}a^{-\frac{5}{2}}\right)\right]\,.
\end{equation}

\begin{figure}[ht]
\begin{center}
\includegraphics*[scale=0.6]{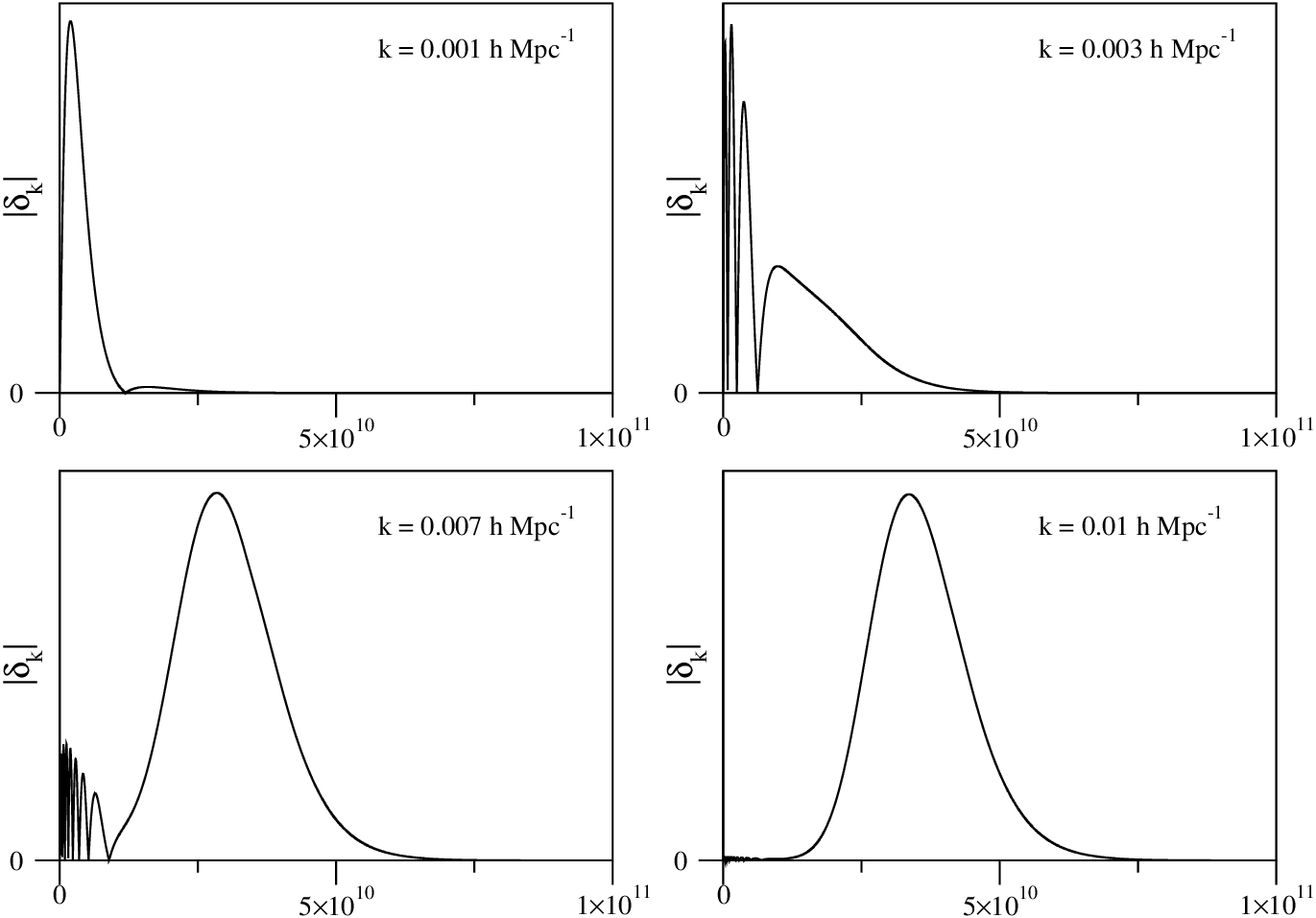}
\end{center}
\caption{Graphs presenting the behaviour of the modulus of the density contrast, $\left|\delta_k\right|$, with the time $t$ (in years), for some values of the wavenumber of the 
perturbation, $k$, for $\gamma=5/3$. Notice the presence of oscillations in the early times, along with the formation of a peak. The vertical scale is arbitrary in all graphs. Also, in all graphs the integral path used was $\Delta t=10^2$ years.}
\label{gamma53}
\end{figure}

As numerical example, we choose the value $\gamma=5/3$ and, following the procedures presented for $\gamma=4/3$, we obtained the graphs presented in Fig. \ref{gamma53}, which represent well the kinds of behaviours we encountered in our computations for several values of $\gamma$.

We also made, now for $\gamma=2/3$, a numerical calculation of the time $t_{peak}$ in which the maximum value of the peak is attained for each mode with wavenumber $k$, and such calculations are presented in Fig. \ref{tpeak}.


\section{Final remarks}

In this work, we present the mathematical analysis of the evolution of 
density perturbations in a flat cosmological model with a modified 
Chaplygin gas acting as a single component, in order to verify whether 
this kind of analysis can be used to distinguish such class of models 
from another cosmological models which have the same solutions for the 
general evolution of the scale factor of the universe. For example, the 
solution
\begin{equation}
a\left(t\right)=\lambda\left(e^{\alpha t}-1\right)^{1/2}\,,
\end{equation}
shown here, can also be obtained in models with a constant bulk 
viscosity or with a decaying $\Lambda$-term of the type $\Lambda=\sigma 
H$ \cite{EueMartin}. Our goal was to verify to what degree such 
degeneracy between models could be broken from the theory alone, without 
resorting to observational data. And our work shows that the analysis of 
the evolution of perturbations can be sufficient, since it produces 
results which can characterize uniquely a model such as the modified 
Chaplygin gas. As an example of this, we can point that a preliminary 
analysis of the evolution of perturbations in the $\Lambda$-decaying 
model for the mode $k=0$ \cite{Saulopersonal} indicates a result 
completely diferent from the one given by equation (\ref{eqz}).

It is also important to notice that we have shown it is possible to find 
a complete analytical solution for the evolution of the density 
contrast, involving spheroidal functions and the conformal time $\eta$, 
{\it in the particular model studied by us}, i.e., a modified Chaplygin 
gas model described by the equation of state,
\begin{equation}
p=\left(\gamma-1\right)\rho-M\rho^{-\mu}\,,
\end{equation}
with $\gamma=4/3$ and $\mu=-1/2$. It is very unlikely that other classes 
of models present the same kind of solution and, as a matter of fact, it 
is very unlikely also that other Chaplygin gas models, with values of 
$\gamma$ and $\mu$ different from the ones chosen by us, will lead to 
analytical solutions. However, it is certain that the analytical 
solutions we found, although limited to a specific choice of parameters 
in a model, can be useful to test numerical codes, to provide some 
insight for more generic numerical solutions, and by themselves are 
interesting for pedagogical purposes.

Another relevant point to be noticed in our work is that our numerical 
analysis, together with the analytical approximate solutions obtained, 
shows that the density perturbation has peaks in its evolution. 
Furthermore, the time in which the peak is maximum depends on the value 
of the wavenumber $k$. Also, the perturbation magnitude is very 
sensitive to the value of $k$. These properties may characterize the 
modified Chaplygin model and help to distinguish it from other models.

To finish, we must emphasize that a thorough analysis of the evolution 
of perturbations in a particular model must include the power spectrum, 
in order to allow a fair evaluation of the possible validity of the 
model. Here, our intention was to show that the results obtained for a 
specific choice of the parameters of the modified Chaplygin gas are 
enough to allow the distinction from another models, from a purely 
mathematical point of view, i.e., {\it even without} the construction of 
the power spectrum. Therefore, a comparison of the power spectrum 
predicted by this model and the one observed remains as a point to be 
attacked in future works.

\section*{Acknowledgements}

S.S. Costa thanks the Brazilian agency FAPEMAT (Funda\c c\~ao de Amparo 
\`a Pesquisa do Estado de Mato Grosso) for financial support during part 
of the time of realization of this work; A.F. Santos thanks the 
Brazilian agency CAPES (Coordena\c c\~ao de Aperfei\c coamento de 
Pessoal de N\'\i vel Superior). S.S. Costa also thanks Profs. Adellane 
A. Sousa, Rosangela B. Pereira and the staff of the UFMT at Pontal do 
Araguaia for their hospitality. Finally, the authors thank Prof. Gustavo M. Dalpian, from UFABC, for his help with computational facilities.

\appendix

\section{\label{ap1}Other forms of the equation for perturbations}

An interesting alternative form of writing the differential equation 
which describes the evolution of the perturbations in the general 
relativistic context, eq. (\ref{Pad1}), is obtained using the scale factor $a$ as 
the variable, {\it i.e.} (cf. equation 4.122 from Padmanabhan \cite{Padmanabhan}),
\begin{eqnarray}
\frac{d^2\delta}{da^2}+\frac{1}{2a}\frac{d\delta}{da}\left(3-15\omega+6v^2\right)
-\frac{3\delta}{2a^2}\left(1-6v^2-3\omega^2+8\omega\right)=-\frac{k^2v^2}{H^2a^4}\delta\,.
\end{eqnarray}
Such result can also be written as
\begin{eqnarray}
\frac{d^2\delta}{d\eta^2}+\mathcal{H}\left[1-3\left(2\omega-v^2\right)\right]\frac{d\delta}{d\eta}
-\mathcal{H}^2\left(1-6v^2-3\omega^2+8\omega\right)\frac{3\delta}{2}=-k^2v^2\delta\,,
\end{eqnarray}
where $\eta$ is the conformal time defined by the relation $dt=ad\eta$, and 
\begin{equation}
\mathcal{H}=\frac{1}{a}\frac{da}{d\eta}\,.
\end{equation}
For classical models of a single fluid in a flat space, with $\gamma\neq 2/3$, one has
\begin{equation}
a\left(\eta\right)=a_0\left(\frac{\eta}{\eta_0}\right)^{\frac{2}{3\gamma-2}}\,,
\end{equation} 
where $\eta_0$ is a constant, and
\begin{equation}
\mathcal{H}\left(\eta\right)=\frac{2}{3\gamma-2}\frac{1}{\eta}\,,
\end{equation} 
so that the equation obeyed by the perturbations is
\begin{equation}
\frac{d^2\delta}{d\eta^2}-\left(\frac{\gamma-\frac{4}{3}}{\gamma-\frac23}\right)\frac{2}{\eta}\frac{d\delta}{d\eta}+\left(\frac{\gamma-2}{\gamma-\frac{2}{3}}\right)\frac{2\delta}{\eta^2}=-k^2\left(\gamma-1\right)\delta\,,
\end{equation}
from where it is easy to see the simplicity of three particular cases 
($\gamma=1,4/3$ and $2$) and which has a general solution in terms of 
Bessel functions,
\begin{equation}
\delta\left(\eta\right)=\eta^{\frac{1}{2}\frac{9\gamma-10}{3\gamma-2}}J_{\pm\frac12\frac{3\gamma+2}{3\gamma-2}}\left(\eta k\sqrt{\gamma-1}\right)\,.
\end{equation}
On the other hand, using $a$ as variable one has the equation
\begin{eqnarray}
\frac{d^2\delta}{da^2}-\frac{9}{2a}\left(\gamma-\frac43\right)\frac{d\delta}{da}+\left(\gamma-2\right)\left(\gamma-\frac23\right)\frac{9\delta}{2a^2}=
-\frac{k^2}{a^{4-3\gamma}}\left(\frac{3}{8\pi\rho_0a_0^{3\gamma}}\right)\left(\gamma-1\right)\delta\,,
\end{eqnarray}
which also can be solved with Bessel functions,
\begin{equation}
\delta\left(a\right)=a^{\frac{9\gamma-10}{4}}J_{\pm\frac12\frac{3\gamma+2}{3\gamma-2}}\left[\frac{3k}{3\gamma-2}\left(\frac{\gamma-1}{6\pi\rho_0a_0^{3\gamma}}\right)^{\frac{1}{2}}a^{\frac{3\gamma-2}{2}}\right]\,,
\end{equation}
and from where it is easy also to see the simplification which can be 
obtained for some particular values of $\gamma$.

\end{document}